\documentclass[manuscript]{aastex}
\usepackage{psfig}

\newcommand{\lta}{$\; \buildrel < \over \sim \;$}
\newcommand{\simlt}{\lower.5ex\hbox{\lta}}
\newcommand{\gta}{$\; \buildrel > \over \sim \;$}
\newcommand{\simgt}{\lower.5ex\hbox{\gta}}

\slugcomment{Submitted to {\it AJ}}
\shortauthors{Howell et al.}
\shorttitle{EUV \& IR Observations of HU Aqr}

\begin{document}

\title{Simultaneous EUV and IR Observations \\ of the Eclipsing Polar HU Aqr}

\author{Steve B. Howell}
\affil{Astrophysics Group, Planetary Science Institute, Tucson, AZ  85705}
\email{howell@psi.edu}
 
\author{David R. Ciardi}
\affil{Department of Astronomy, University of Florida, Gainesville, FL 32611}
\email{ciardi@astro.ufl.edu}

\author{Martin M. Sirk}
\affil{Space Sciences Laboratory, UC Berkeley, CA 94720}
\email{sirk@ssl.berkeley.edu}

\author{Axel D. Schwope}
\affil{Astrophysikalisches Institut Potsdam, An der Sternwarte 16, D-14482
Postdam, Germany}
\email{aschwope@aip.de}

\begin{abstract}
We present simultaneous EUV and infrared (J,K) observations of the 
polar HU Aqr obtained
during August 1998 when the star was in a high mass accretion state. 
EUV and IR light curves and EUV spectra are presented and compared with
previous observations.
The accretion region on the white dwarf has increased
in temperature (124,000K to 240,000K) and radius (0.04 R$_{WD}$ to 0.06 R$_{WD}$) 
compared with previous EUV observations made
during low mass accretion states. 
The EUV and IR
photometric observations are shown to have a similar appearance as a function of
orbital phase. The EUV photometry shows rapid changes and provides evidence for
mass accretion via blobs. The high state IR light curves present an
asymmetric double-humped 
shape with J=14.8 and K=14.1. We applied an ellipsoidal model fit to the 
observations and the result indicates that the cause of the modulated 
shape is both due to ellipsoidal variations from the Roche Lobe filling 
secondary star
and a complex flux combination 
probably dominated at all orbital phases by cyclotron emission. 
The source of maximum cyclotron
emission appears to be in the accretion column 
high above the white dwarf surface.
\end{abstract}


\section{Introduction}

HU Aqr (RX J2107.9--0518) 
is a member of the polar or AM Her type of cataclysmic variable.
Polars consist of an interacting pair of stars (white
dwarf + red dwarf) in which the white dwarf primary is strongly magnetic.
The material accreted from the low mass secondary is magnetically
controlled over the final portion of its flow being forced to funnel along
the field lines and impact directly onto the white dwarf surface. The general 
location of the material impact site on the white dwarf surface is termed the 
accretion region.
HU Aqr has an orbital period of 2.08 hr and the white dwarf magnetic 
field strength is estimated to be 36 MG. 

High
energy observations of polars provide us with direct information related
to the accretion regions near the white dwarf surface surrounding the
magnetic poles (See Sirk \& Howell 1998; Schwope et al. 2001a). 
A number of high energy observations have been obtained for HU Aqr due
to both its brightness at X-ray and EUV wavelengths and the fact that it
is an eclipsing system. Eclipsing systems allow absolute orbital phase
information to be obtained without ambiguity caused by accretion stream
or accretion region eclipses. Phase dependent phenomena are then able to be
referenced to an unchanging fiducial. 
A review of the observational history of HU Aqr, in
particular the high energy observations, is provided by Schwope
et al. (2001a). 

Infrared observations of cataclysmic variables are useful as a tool to
understand the secondary star which is typically of late spectral type and bright in
the IR. These parameters are addressed through observations
of ellipsoidal variations and the secondary star
spectral energy distribution. IR observations also provide information
about the cooler areas of the
accretion stream and cyclotron radiation produced in the hot regions near the
accretion spot. 
The usefulness of IR photometry is illustrated
in Ciardi et al. (1998) and Howell et al. (2001) and 
IR spectroscopy in Howell et al. (2000).

In this paper, we present simultaneous EUV and infrared observations for
HU Aqr obtained during a high mass accretion state in August 1998. The
EUV observations allow us to measure the size and location of the
accretion region and its temperature while IR photometry presents
contributions from the accretion stream and coupling region, 
the secondary star, and cyclotron radiation. The only previous attempt at such
a multi-wavelength simultaneous approach was by Watson et al. (1989)
using EXOSAT X-ray, optical, and IR data for the polar EF Eri. 
These authors concluded that the
coincident dips seen in all the bands were caused by absorption
in the accretion stream as it crosses the line of site to the accretion pole.
Our new data allows a more detailed understanding of the process and we find
that the two apparently
distinct wavelength bands seem to have previously undiscovered
commonalities as we find a correlation in their phase-resolved flux
distributions. 

\section{Observations}

\subsection{{\it EUV} Observations}

The {\it EUVE} satellite performed simultaneous spectroscopic and
photometric observations in the EUV spectral range (Bowyer and Malina
1991) The principle instrument on board consists of a telescope which
contains an imager and three separate spectrographs covering the 
range of $70-750$\AA. The bandpass of the deep survey (DS) imager is set
by the Lexan/Boron filter, with a maximum transmission at $91$ \AA~with
a $90\%$ bandpass of $67-178$ \AA. The imager allows for collection of
photometric data simultaneously with the spectroscopic data.
The {\it EUVE} obtains short wavelength (SW), medium
wavelength (MW), and long wavelength (LW) data as three separately imaged
dispersions covering the ranges of $70-170$\AA, $150-350$\AA, and
$300-700$\AA~respectively. All collected photons are position and time
tagged providing high time-resolution and allowing the production of
detailed light curves (e.g., Sirk and Howell 1998) and spectra (e.g., Craig
et al. 1997; Mauche 1998; \& Howell et al. 1997). Details of the
photometric properties of the imaging telescopes on board the {\it EUVE}
may be found in Sirk et al. (1997) and the spectroscopic instruments
are reviewed in Abbott et al. (1996).

The {\it EUVE} observations of HU Aqr were obtained during the time
period 1998 August 27 (21:47:10.0 GMT) to 1998 August 29 (21:30:49.0 GMT)
with a total on-source integration time of 65.432 ksec. This time
coverage contained 8.7 consecutive orbital periods of HU Aqr. During the
{\it EUVE} observation, HU Aqr was detected with a mean count
rate of 0.8 counts sec$^{-1}$ photometrically and $\sim$0.1 counts sec$^{-1}$
(near 100\AA) spectroscopically. No EUV flux was detected long-ward of
$\sim$110 \AA~due to absorption by the ISM. 
The high count rate observed made HU Aqr approximately
15 times brighter than any previous observation made with {\it EUVE}
(see below) which spanned the time period of 1996-1997. The {\it EUVE}
spectral data were extracted and reduced to phased-resolved 2-D images
as described in the {\it EUVE} users manual, and then to 1-D spectra as
discussed in Hurwitz et al. (1996). The photometric data reduction
proceeded as described in Howell et al. (1995).

Figure 1 presents the mean EUV light curve obtained during the August 1998
observation binned in 15 sec intervals. 
Various phases of interest, which will be used later on,
are identified on the figure and we note that the eclipse egress lasts
less than 2 seconds. All data in this paper are phased on the ephemeris
of Schwope et al. (1998),
$$ T(HJD) = 2449217.345162(27) + 0.08682041520(82)N $$ 
with phase 0.0 representing the time of inferior conjunction of the
secondary star. This ephemeris is equivalent to that of Schwope et
al. (2001a) to within 10$^{-4}$ seconds.

The previous observations of HU Aqr made with {\it EUVE}
had too low of a count rate for useful spectra to be extracted. With the
higher flux present in the current dataset, we are able to produce not
only a summed spectrum covering the entire on-source time, but single
spectra from phases of interest within the orbital cycle. Figure 2 shows
our results for HU Aqr, labeled as to their
phases as marked in Figure 1 and binned to a final spectral resolution of 
0.8\AA. The 
spectra shown in Figure 2 have been corrected for {\it EUVE} deadtime,
primbsching, and the spectrograph effective area response function 
(See Abbott et al. 1996). 

\subsection{Infrared Observations}

Using the 2.34 m infrared telescope at the Wyoming Infrared Observatory
(WIRO) near Woods Landing, WY, HU Aqr was observed on UT 1998 August 29
\& 30 in the near--infrared broadband filters J \& K using the Aerospace
Corporation 256 $\times$ 256 NICMOS3 camera. The camera has a spatial
resolution of 0\farcs43 per pixel for a total field of view of
110\arcsec\ $\times$ 110\arcsec. The array has a quantum efficiency of
$\gtrsim$ 60\% over the $1-2$ \micron\ wavelength range, an approximate
dark current of $\lesssim$ 1~e$^-$/second, and a readout noise of 
$\lesssim$ 30 electrons/pixel.
 
To minimize the effects of a variable sky background, the data were
acquired in image ``nod pairs.'' A nod pair consists of a ``$+$beam''
image and a ``$-$beam'' image which are spatially separated from each
other by 20\arcsec\ in declination. The integration time of a single
frame within a nod pair was 20 seconds. The image collection was
performed such that a pair of images (one per ``beam'') was taken
before changing filters. The images for the ``nod pairs'' are thus
very close in time, $\sim$11 sec, which incorporates the readout and
writing to disk of the first image and the slewing of the telescope by
20\arcsec\ (including settling time). The nod of the telescope was
chosen carefully so that the source and two non--variable comparison stars
remained within the field of view for both beams. After each filter--nod
sequence, the filter was changed and the sequence repeated. Succesive
filter pairs are separated by $\sim$2.5 minutes.

To increase the signal-to-noise ratio of the photometry, images within a
``nod--pair'' were registered and co-added, and standard differential
photometry was performed on each co-added image as described in Howell,
Mitchell, \& Warnock (1988). Absolute photometry was obtained by
observing a near--infrared standard star from the list of Elias et al. 
(1982) just prior
to the start of the HU Aqr observations. The HU Aqr observations for 
the two nights
were combined and phased on the orbital period using the same
ephemeris as the {\it EUVE} data. The final J and K light curves are
shown in Figure 3 where they are compared with the 
1996 J and K light curves from
Ciardi et al. (1998). HU Aqr had mean J and K magnitudes of 16 and 15 respectively
in 1996, while our new 1998 measurements have mean J and K 
magnitudes
of 14.8 and 14.1 respectively.

\section{Results}

\subsection{EUV}

Figure 4 presents all the {\it EUVE} observations of HU Aqr.
The August 1998 observation (bottom panel) has a
much greater count rate than any of the other light curves in Fig. 4 and its
shape has dramatically changed. The time sequence seen in
Fig. 4 prior to the 1998 observation
was originally thought to represent a progression of HU Aqr into
a low mass accretion state. However, the 1998 observation would seem to
indicate that {\it all} the previous {\it EUVE} observations were during
a low mass accretion state, some just lower than others. Perhaps the most dramatic
change observed in the 1998 EUV photometry of HU Aqr is the fact that
the post-eclipse flux is now greater than the pre-eclipse flux, the
opposite behavior to essentially all the low mass accretion state observations 
(Fig. 4). {\it ROSAT} observations of HU Aqr made in 1993 show 
similar count rate ratios compared with the August 1998
{\it EUVE} observation and also have higher 
post-eclipse fluxes (See Fig. 2 in Schwope et al.
2001a).

Many {\it EUVE} and {\it ROSAT} observations contain no flux from phase
0.9 to 0.97. The reason for this lack of flux is probably due to the
accretion curtain blocking our view of the accretion region
during this phase interval (See Sirk \& Howell 1998; Schwope et al.
2001a). Additionally, we will see below that the flux level within the
0.9 to 0.97 phase window can be highly time variable. 

Figure 5 presents the eight consecutive EUV HU Aqr light curves obtained 
during the August
1998 observation. 
Each of the eight panels spans about a quarter day or 3 
binary orbits of HU Aqr.
The dotted line in Fig. 5
is the mean light curve of Fig. 1 illustrating that changes from orbit
to orbit clearly exist. Note, for example, the change in flux near phase
0.925 related to effects in the
far field accretion stream and the accretion curtain (near field stream) 
and the fast changes in absorption near phase 0.75 due to the near field
accretion column. Sirk \& Howell (1998) and Schwope et al. (2001a)
show that changes in the shape of the soft X-ray/EUV light curves occur from
observation to observation (months to years apart) and these 
changes provide evidence 
for the movement of the accretion stream due to changes in
the mass accretion rate.
Figure 5 provides
evidence that rapid (few hour) changes occur during this high mass accretion state
and are 
probably related to local opacity variations or blobs within the accretion 
stream and column (See \S4).

Figure 6 shows our August 1998 spectrum during the bright phase (bottom
panel) compared with the spectral sum of three previous observations
(made during 1996) co-added over all phases with non-zero
EUV flux. The final binned spectral resolution is 0.54\AA~for 
both plots in Fig. 6. 
While the older summed spectrum is of lower S/N, it is clear
from its appearence that it is produced by a cooler accretion region
than that in the August 1998 observations (see below).

\subsection{Infrared}

Infrared J and K light curves from 1998 and 1996 are shown in Figure
3. A comparison of the the light curves from the two epochs reveals some
intriguing features. The most striking diffference
between the light curves is the 1998 data are, on average, 2.5--3 times
brighter than 1996 data. The infrared brightness increase is $\sim$15 times
lower than the EUVE brightness increase.

At first glance the 1998 and the 1996 light curves appear to be 
similar in structure. Both sets of light curves display the deep stellar
eclipse bottoming out at J$\sim$16.4 mag and K$\sim$15.4 mag.
While the overall infrared brightness has increased between
1996 and 1998, the depth of the eclipse has not changed indicating that
the infrared eclipses are total and the bottom of eclipse is
representative of the back of the secondary star (See Ciardi et al.
1998).

In addition to the deep stellar eclipse, both J and K light curves are
double-humped. The 1996 observations show two
symmetric humps centered at orbital phases 0.25 and 0.75 which are 
well fit with an ellipsoidal variation model of the secondary stellar
photosphere. The ellipsoidal model from Ciardi et al. (1998) is shown in
Figure 3 and is overplotted upon the 1996 and 1998 data. The model,
which explains the global variation of the 1996 data, does not explain
the variations observed in the 1998 data. The 1998 ``humps'' are
out of phase with those expected for ellipsoidal variations and are not 
symmetric in shape or amplitude. 
The ellipsoidal variations of the secondary star probably did not disappear
between 1996 and 1998 (cf. Howell et al. 2000), 
but rather during this high accretion state, 
the secondary star flux is now 
over powered by orbitally 
modulated variations from another source. We note here a
general warning to observers of polars that infrared light curves 
with ``double--humped''
structures do not necessarily consistute ellipsoidal variations of the
secondary star. This same misconception has been seen in the optical as well
(Howell et al. 2001). Thus, while ellipsoidal variations may still effect the
light curve shape by providing some amount of the modulated structure,
we will see below that the IR light curve in HU Aqr is quite 
complex and dominated by another source.

\section{Discussion}

\subsection{EUV: Light Curves and Spectra}

Sirk \& Howell (1998) developed a three-dimensional model which allowed
EUV photometric data to be fit in terms of various parameters dealing
with the size, shape, and location of the accretion region on the white
dwarf surface. Table 1 summarizes the findings from their original work
on the 1996 {\it EUVE} observations of HU Aqr along with our new values
determined from model fits to the current data. It has been assumed that the
orbital period of HU Aqr has reminained constant and we have used the new
determination of the system inclination, 85.6 degrees (Schwope et al.
2001a), instead of the value of 81 degrees available to Sirk \& Howell.
Schwope et al. (2001a) determined the size of the accretion region in HU Aqr
for the high state 1993 {\it ROSAT PSPC} observations by 
analyzing eclipse ingress timings 
and by applying the 3-D Sirk and Howell model.
In both cases, they
found a consistant size of 0.052R$_{WD}$ (1$\sigma$=0.02). 
This value is essentially the same as our high state determination listed in Table 1.
In addition, there are enough photons to examine eclipse egress
in the 1998 observations and its duration of 1.5-2.0 seconds, sets an upper limit
on the EUV emitting region of 0.07 R$_{WD}$.
Our Table 1 values for the 
1996 and 1998 spot heights are the same but they are 
larger than the value of 0.014 R$_{WD}$ given
in Schwope et al.
No uncertainty is given for the value of the height determined by these authors, but
if it is similar to our uncertainty ($\pm$0.003), then we are not too far apart in 
our calculated heights.
We find that the accretion region in HU
Aqr has increased its area by a
factor of $\sim$2.8, if approximately spherical in cross
section, during this high mass accretion state. 
Uncertainties in the accretion region modeling procedures are fully discussed
in Sirk \& Howell (1998).

The three large dips seen in the EUV light curve (Fig. 1) are due to
local absorption of EUV flux by the near field and far field accretion
columns caused by our line of sight through to the accretion region.
Sirk \& Howell (1998) found that a comparison of the EUV spectra
obtained for UZ For, VV Pup, and AM Her during their broad dip phases
with that observed during their bright phase (see their Figure 9) led to
the result that spectra observed during the broad dips were softer
than those obtained during the (assumed) unocculted bright phase. To see
if this result holds for HU Aqr as well, we 
have taken the sum of our two HU Aqr ``dip" spectra (Dip1 \& Dip2;
bottom two panels in Fig. 2) and compared them with the Bright phase
spectrum. The low S/N in the dip spectrum made the comparison rather noisy but
over the region of $70-90$\AA, the spectrum 
appeared softer than that collected during the
Bright phase, exactly what was found for UZ For, AM Her, and VV Pup.

Sirk and Howell
(1998) have shown that the broad dip is caused by material very close
to the accretion region (the near field stream) 
and modeled it with a cylindrically symmetric,
uniformly dense absorber immediately above the accretion spot.
Schwope et al. (2001a) show that it is not a cold absorber which is responsible
for the broad dip.
The broad dip in the EUV light curves of HU Aqr varies on short (binary period)
time scales
as can be seen in Figure 5. The large
variations from orbit to orbit indicate that the broad dip
is a changable feature in both phase and shape: the ``broad dip'' shape
becoming manifest only when many individual orbits are averaged together.
Warren, Sirk, \& Vallerga (1995) found a similar behavior in the polar UZ For.

The many {\it ROSAT} and {\it EUVE} datasets for HU Aqr were used to
search for correlations between accretion
rate, spot latitude, stream dip phase, and broad dip phase as a function of time. 
The results are presented in Table 3 in Schwope et al (2001a).  Most of the
observations show the broad dip to occur at early orbital 
phases ($\phi = $ 0.69 to 0.77),
however, a third of the observations show no clearly defined broad dip at all.
To graphically illustrate the changes in the broad dip in HU Aqr, we show a
normalized average EUV light curve (dotted line) drawn on each of the individual
light curves in Figure 4. This average curve was produced by 
taking every {\it EUVE} light curve of HU Aqr, normalizing each to its
maximum value, averaging them all together, and then over-plotting them (scaled by
each light curve maximum) on each single
epoch light curve in Figure 4.
Comparison of the ``dipless" individual light curves in Figure 4 
(May, July, \& September 1996) with the normalized 
over-plotted average light curve reveals 
a deficit of flux at later phases 
(around $\phi = $ 1.0 to 1.1).
An extreme example of this effect can be seen in
the {\it ROSAT} HRI April 1996 observation (Figure 2, Schwope et al. 2001a)
where the second half of the light curve is nearly absent.

A quantitative correlation between the broad dip phase and the mass accretion rate
is difficult to access but the following appears to be true. 
At a high accretion rate where we see the broad dip occur
at an early phase, the low density portion of the ballistic stream latches
onto the magnetic field first and strikes the WD at a relatively low
longitude, and the higher density portion of the stream (blobs) penetrate
further into the magnetic field and land on the WD at a greater longitude.
Our view to the accretion spot through the column will then be most obscured
at early phases (i.e., causing a flux deficit).

For the broad dip to occur at an early phase, it requires absorbing material to
lead the EUV accretion spot in longitude on the white dwarf surface.
When at later phases, the material must lag in longitude.
We conclude that the near field accretion curtain 
is non-uniform in density at any given phase at any given time.
Thus, at different times we are looking through varying amounts of absorbing 
material and
these differences in column density are probably influenced by how
the ballistic stream attaches on to the magnetic field lines within the coupling
region. Even small mass accretion rate changes or blobby accretion 
will supply varying ram
pressure causing the field lines to bend and changing the exact location of the
coupling region for any given field line.

Mauche (1998) provided a detailed look at modeling the EUV spectra of
polars. He concluded that absorbed blackbody models provide the best
phenomenological description for the $70-180$\AA~spectra of polars. However,
Mauche and prior work by Paerels et al. (1994) both note problems with such a
simple approach. The weak absorption edges and lines (mostly due to Ne species)
are not properly dealt with and blackbody fits are unable to produce the observed
EUV fluxes. Use of an irradiated solar composition stellar atmosphere model is 
likely to be more proper but models of this type need to be improved by adding in
non-LTE effects, the underlying white dwarf, and absorption lines. 
Likewise, the quality of the spectral data to be modeled must
greatly improve in order to allow quantitative analysis.

An additional complexity pointed out by Paerels et al. (1994) is the fact that
a change in the fitted blackbody model continuum level will cause 
corresponding changes in the observed absorption line strengths. The use of 
simple blackbody models (especially to fit a mean spectrum) forces a ``best fit
continuum" criteria on the user, thereby fixing the apparent line strengths.
Flucuations in intensity and spectral distribution which occur over time and
orbitalphase are not accounted for. The S/N of the spectrum also plays a role here
as it influences the selection of the best fit. A detailed example is given in 
Paerels et al. 
Given the moderate S/N of our HU Aqr spectra and the fact that an absorbed
blackbody model does provide a decent fit to the observations, 
we opt to model our data in
this manner. However, the reader should keep the above caveats in mind.

Using
point sources (white dwarfs and B stars) observed with {\it EUVE} for
the purpose of mapping out the ISM, we can estimate the interstellar
column to HU Aqr. Ten sources in the {\it EUVE} data archive located in the
direction of, and near the distance of HU Aqr (125 pc; Ciardi et al.
1998) were used to provide an initial estimate for the column density to HU Aqr 
of log
N$_{H}$=19.75 (assuming HeI/H=0.1 and HeII/HeI=0.01). Starting with this
column density estimate and allowing our blackbody temperature and log
N$_{H}$ to be free parameters, we determined a best fit absorbed blackbody
solution for our summed spectrum (bottom panel, Fig. 6). Our best
fit yields a temperature of 240,000$\pm$40,000 Kelvins (20.68 $k$T (eV)) 
for the accretion region with a column to HU Aqr of log
N$_{H}$=19.48$\pm$0.32. These values are in excellant agreement
with those determined by Schwope et al. (2001a) for 
HU Aqr during 
a similar high mass accretion state {\it ROSAT-PSPC} observation 
made in October 1993. 
Attempts to fit absorbed blackbody
models to the individual spectra from our August 1998
data or to the summed spectra from the 1996 datasets yielded large
uncertainties due to their low counting statistics. 
The accretion region temperature for the 1996 low mass accretion state
was found to be much cooler, with a best fit near $\sim$124,000 K, based
on the spectral slope and assuming that the column density remained
constant.

Spectral lines or line edges
have been observed for a few polars in {\it EUVE} spectra (See a review
by Mauche 1998).
Mauche (1998) presents a detailed re-analysis of
these same data and finds convincing evidence in a subset of the polars
for lines due to Ne VI, VII, and VIII. 
Our August 1998 HU Aqr summed spectrum reveals a few features
that may be real atomic absorption absorption lines and is
of sufficient S/N to allow a qualitative examination of
these possible spectral features. Searching the likely line species for polars
present in the $70-100$\AA~range (N, O, Ne, Mg, and S),
absorption lines due to Ne VIII at
73.5, 76(?), and 98.2 \AA~are the only set of consistant
and possibly believable features. Ne VI line edges at 78.5 and 
85.2 \AA~and that for O VI at 89.8 \AA~may be present. 
Ne VIII (98.2 \AA)
seems to be present in the 1996 low state spectrum as well.

The ionization potential for Ne VIII (239 eV) is too
high to be provided by the EUV blackbody emission as the accretion region
temperature (derived above) only provides $\bar{E} = \frac{3}{2} k T =
31$ eV. A temperature near 20 million Kelvins (20 keV) is necessary to
begin to excite these Ne inner shell transitions. Thus, if we are to believe
the Ne VIII spectral features are real, we must invoke an additional
heating source in the white dwarf atmosphere at or near the accretion
region. 

Ramsey et al. (1994) found that essentially all polar spectral energy
distributions in the hard and soft (EUV) X-ray region are best fit by a
two temperature model (an absorbed few 100,000K blackbody plus a harder
thermal bremsstrahlung component). Typical shock temperatures (thermal
bremsstrahlung fits) derived for X-ray observations of polars were found
to be of order 15--30 keV, quite sufficient to produce Ne VIII. Schwope et
al. (2001a) modeled HU Aqr with a bremsstrahlung component having a 
temperature of 20
keV, consistant with Ramsey et al. and sufficient to produce the neon lines. Thus
for HU Aqr, Ne VIII absorption features superimposed on a blackbody
spectrum would be consistant with a typical two-temperature model for
the high energy emission from the accretion region in a polar. The
existence of a harder component is believed to indicate that there is
significant external heating of the white dwarf atmosphere by the
bremsstrahlung radiation (an irradiated atmosphere) at and near the
accretion region. The lack of any visible O VI lines in the HU Aqr
spectrum suggests that the white dwarf atmosphere is heated 
only to small depths near
the accretion region (van Teeseling et al. 1994
\& See Fig. 1 in Paerels et al. 1996).

Polars are well known to show a ``Soft X-ray Excess" (Ramsay et al. 1994; Warren
and Mukai 1996), that is, the ratio 
L$_{EUV}$/L$_{X-ray}$ is greater than the value of 0.55 predicted from 
theory (King and Watson 1987). Given that we have (non-simultaneous) 
EUV and X-ray 
observations of HU Aqr in both low and high states let us determine its soft X-ray
excess during these times. From our 1998 high accretion state and 1996 low
accretion state EUV observations we find L$_{EUV}^{high}$=1.16 $\times$ 10$^{32}$
ergs sec$^{-1}$ and 
L$_{EUV}^{low}$$\simeq$3.5 $\times$ 10$^{31}$ ergs sec$^{-1}$.
Using the X-ray flux observed during the 
similar 1993 high accretion state seen by {\it ROSAT} (Schwope et al. 2001b),
we find L$_{X-ray}^{high}$=2.0 $\times$ 10$^{31}$ ergs sec$^{-1}$. 
Low mass accretion states (such as during 1996-97) have X-ray fluxes 
which are about 20 times lower overall than during a high state, thus 
L$_{X-ray}^{low}$$\simeq$1.0 $\times$ 10$^{30}$ ergs sec$^{-1}$.
Schwope et al. (2001a) showed that during the accretion region 
eclipse, HU Aqr was detected with L$_{X-ray}$=2.2 $\times$ 10$^{29}$ ergs
sec$^{-1}$ ($\sim$0.2 L$_{X-ray}^{low}$) 
apparently due to chromospheric activity on the secondary star.
Using these determinations of the high energy luminosity of HU Aqr, 
we find L$_{EUV}^{high}$/L$_{X-ray}^{high}$=5.8 and
L$_{EUV}^{low}$/L$_{X-ray}^{low}$$\geq$35.

Both of our luminosity ratios are within the range determined 
for HU Aqr by Ramsay et al. (1994) of 3.1 (with a range 
of 1.5-33.4)
during an apparent 1992 low mass accretion state (L$_{X-ray}$=1.9 $\times$
10$^{30}$ ergs sec$^{-1}$). While it appears that the soft X-ray excess
in HU Aqr is greater during low mass accretion states, the range presented
by Ramsay et al. is typical of what one finds within the uncertainties of
model fitting with the low mass accretion states being of higher uncertainty due
to their lower signal. If the deposition of mass blobs below the white dwarf
surface is the cause of the soft X-ray excess as presently believed (See King
1995), our results may indicate that accretion by 
dense mass blobs during times
of lower \.M occur in a larger proportion compared with accretion of 
lower density gas.

\subsection{Infrared: Light Curves}

It is reasonable to assume that the increase in infrared flux from 1996 to 1998
is directly
related to the increased mass accretion state in HU Aqr 
as evidenced by the much brighter EUV flux. 
The EUV flux increase is both a
result of the thermal increase and the size increase 
of the accretion region on the surface of the white dwarf (See Table 1). 
But where does 
the excess infrared emission emanate from? 
The three mostly likely sources are the
thermal emission from the accretion region, cyclotron emission from the
accretion region and column, and thermal emission from the accretion stream.

In Figure 7, the secondary star contribution to the infrared
light curves has been subtracted away from the 1998 observations 
by scaling the ellipsoidal variation
model from Ciardi et al. (1998) to the bottom of the stellar eclipse.
The ellipsoidal subtracted infrared
light curves still show modulations spanning nearly an order of magnitude and of a
complex nature and no longer double-humped.
Interestingly, the strongest excess infrared emission occurs when the accretion region
is self-eclipsed by the white dwarf (region 7, Faint phase). This fact alone 
indicates
that not all of the IR emission emanates from the accretion region located on the
surface of the white dwarf, but rather a significant fraction must
come from elsewhere.

The secondary star subtracted J \& K fluxes (Fig. 7) 
and the EUV light curve (Fig. 1) are normalized
to their maximum value and directly compared in Figure 8. Except for
the infrared flux being non--zero during the EUV Faint Phase (region 7),
the light curves are similar in morphology. In region 1 (Rise),
there is a slight increase of infrared flux matching the EUV rise. The
infrared rise is nearly twice as long in duration as the EUV rise probably
a result of the infrared emission emanating from a region more
extended than the EUV emitting region.

The overall decline of the EUV flux across regions 2--4 (the Dip phases)
is matched in general by the infrared light curves. 
However, unlike the EUV, the general infrared flux
decrease across the Dip phases is not a result of the
accretion column passing in front of the emission region, but is likely
the result of a change in the viewing angle of the beamed cyclotron
emission (See Wickramasinghe \& Ferrario 2000). 
Cyclotron emission is at its strongest when viewed at an angle
of 90 degrees to the magnetic field lines which, depending on 
the exact orientation of the
accretion column, corresponds to orbital phases of 0.6--0.7 in HU Aqr. 
It is only in the
Stream Dip (region 4) and possibly in Broad Dip2 (region 3) 
that there is a corresponding dip in the
infrared, and it only appears at J. The $<1$\% dip in J is not as large as the
EUV dip which is total ($>4$\%), but it does imply that the accretion stream
is more optically thick at J than at K. 

The IR ingress during stellar eclipse (region 5) is similar in
length to the ingress observed in the EUV but occurs slightly earlier in phase (See
Schwope et al. 2001b). 
The infrared
egress rises sharply with that of the EUV flux but the total recovery time
is significantly longer than in the EUV.
The difference
in the infrared egress time is likely a result of viewing 
geometry and the
larger IR emitting volume. As the
secondary approaches inferior conjunction, the accretion column is
viewed more and more straight-on. This orientation will also provide decreasing 
infrared cyclotron
emission.  Therefore, at the beginning of stellar eclipse, the
projected area of the accretion column is relatively small and quickly
eclipsed.  But as the eclipse ends, the accretion column has rotated
more perpendicular to the line of sight, increasing the time required to
come fully out of eclipse. 
As the column rotates
to a more perpendicular orientation with the viewing angle, the total 
IR emission increases as we see during the Faint (region 7) phase.
This interpretation is slightly complicated
by the fact that the level of (beamed) cyclotron emission from the 
accretion column is also
dependent upon the line of sight viewing angle.  

It would not be surprising if cyclotron emission dominates the infrared
emission during the phases when the EUV accretion region is in view. However,
can cyclotron emission contribute significantly during other phases, particularly
the EUV Faint
phase (region 7), or will the infrared emission be dominated by thermal
emission from the coupling region and accretion stream? To test these ideas
the J data was re-sampled at the K data rate. Because the J and K
data were obtained in a sequence (K,J,K,J,...), for each K point the two adjacent
J points in time were 
fit with a low order spline and the best-estimate J
value was then interpolated for the exact time of each K data point. 
The intensity ratio of the J
flux to the K flux was determined as a function of orbital phase and is
shown in Figure 9. 

Blackbody radiation, on the Rayleigh-Jeans tail, follows a
$\lambda^{-4}$ distribution which corresponds to a J/K flux ratio 
of $\sim 11$\footnote{As a check on this value, we re-examined the J/K flux 
ratio for HU Aqr during the low
mass accretion state reported on in Ciardi et al. (1998). During those observations,
there was essentially no mass transfer and thus almost no cyclotron emission present.
The J/K flux ratio was $\sim$10, the value expected for 
essentially pure blackbody emission.}. 
Cyclotron radiation, however, has a more complicated wavelength dependence.
At long wavelengths, the ``continuum" consists of optically thick emission
which follows a Rayleigh-Jeans wavelength dependence. For magnetic field strengths
typical of those in most polars,
the optical and infrared spectral regions show the cyclotron continuum 
to be highly modulated by 
harmonic structures called cyclotron humps. 
The spectral dependence of cyclotron emission falls to a much
shallower value (near $\lambda^{-1.5~{\rm to}~-2.4}$) within the regions 
which are modulated by cyclotron humps with the slope becoming steeper as one
moves to earlier harmonics. 

Glenn et al. (1994) observed cyclotron harmonics 4, 5, and
6 in the optical spectrum of HU Aqr 
and fit them with a 10 keV plasma cyclotron model and a white dwarf
magnetic field strength of 36 MG. These results have been confirmed by Schwope et al.
(2001b). Thus,
the first three cyclotron harmonics in HU Aqr 
will modulate the continuum in the J,
H, and K bands as was pointed out by Ciardi et al. (1998).
This interpretation is also consistent with
model cyclotron spectra for a 10 keV plasma and B=35 MG 
calculated by Wickramasinghe \& Ferrario (2000, See their
Figure 32). 
Therefore, using a $\lambda^{-2.4}$ dependence for the IR cyclotron
spectrum in HU Aqr, we would expect a J/K flux ratio of
$\sim 4.3$ if the flux output is dominated by cyclotron emission.

Figure 9 reveals that the flux ratio (F$_{1.2 \mu m}$/F$_{2.2 \mu m}$)
is generally flat throughout the
orbit of the system, never climbing above $\sim$4.8 except
during the stream dip and fall phases (regions 4 and 6).
In the Faint
phase (region 7) the ratio is almost exactly 4.3, making the EUV faint phase
infrared emission consistent with being essentially pure 
cyclotron emission. This is
somewhat surprising given that the white dwarf has self-eclipsed the accretion
region during these phases. However, during this high mass accretion state, 
the accretion column may extend far
enough above the white dwarf surface such that significant cyclotron
emission is visible even during the time when the accretion 
region is self-eclipsed. This is a surprising and unexpected hypothesis.

To see if such a tall accretion column may be possible, 
we calculate the
shock height above the accretion region, that is, the approximate 
upper limit to where the 10 keV electron plasma would be confined. 
Using the mass estimate for the
white dwarf, 0.6-7 M$_{\odot}$ (Schwope et al. 2001b), the mass accretion rate
during an assumed 
similar high mass accretion state, \.M = 6 $\times$ 10$^{-11}$ M$_{\odot}$ yr$^{-1}$ 
(Schwope et al. 2001a), and the expression for the shock height from
Frank et al. (1992, Eq. 6.44), HU Aqr's shock will extend approximately
0.14 R$_{WD}$ above the white dwarf surface.
Geometric arguments show that if the shock height were at least
0.2 R$_{WD}$ it would be visible to an observer even during the EUV Faint phase.
Schwope et al. (2001b) examine the cyclotron radiation from HU Aqr in detail. 
They also propose that the cyclotron emission originates from a large height, 
higher than the soft X-ray emission, but only 0.03 R$_{WD}$. Fischer \& Beuermann
(2001) use 1-D hydrodynamic arguments to derive the
location from which most of the cyclotron radiation emerges, i.e., the shock height. 
Applying their
formulation to HU Aqr, the height of maximum cyclotron emission is calculated to be
near 0.28 R$_{WD}$. This value is two times that determined from the simple
equational form of Frank et al. given above, but is similar to the value needed
herein to allow cyclotron radiation to be observed at all phases.
Thus, given the uncertainties in the values used for these calculations, 
it seems plausible that significant cyclotron emission is observable 
throughout the orbit of HU Aqr during high mass accretion states.

The overall increase in the J/K flux ratio from phase 0.7 to phase 1.1, and its
peak near phase 1.05, is 
likely a result of the {\em relative} increase of the thermal IR emission
from the accretion region during the phases for which the observer has the most
direct view.
Additionally, during these phases the line
of sight is increasingly straight onto the accretion column which
decreases the strength of the (beamed) cyclotron emission.
Thus, while
the overall IR emission is {\em lower} in the phase interval 0.7--1.1
(see Figures 7 and 8), the
relative contribution from thermal IR emission is {\em higher}, causing a
rise in the J/K flux ratio.  However, the ratio never approaches 11 as
expected for pure blackbody emission, indicating that cyclotron emission
remains dominate throughout the orbit.

\section{Conclusion}

We have presented simultaneous EUV and infrared high mass accretion state 
observations of the 
polar HU Aqr. The accretion region on the white dwarf shows an increase
in temperature and radius by $\sim$2 times compared with results obtained 
during a low mass accretion state: the temperature increased from 124,000K to 240,000K
and the radius from 2.2$\times 10^7$ cm to 3.7$\times 10^7$ cm.
A two temperature model consisting of a hot thermal bremsstrahlung component
and an absorbed blackbody component seems to fit the EUV observations.
The EUV and IR
photometric observations are shown to have a correlation
with orbital phase although caused by distinct processes. HU Aqr had mean high state J
and K magnitudes of 14.8 and 14.1 respectively.
We have shown that the high mass accretion state IR 
light curve double-humped structure is {\it not} due to ellipsoidal 
variations from
the secondary star but instead is caused by 
strong geometric modulation of the apparent size of the 
emitting region.
Our results also show that during this high mass accretion state, the IR flux is 
dominated at all orbital phases by cyclotron emission emerging from high above 
the white dwarf surface, near 0.2 R$_{WD}$.

We wish to thank M. Huber for obtaining the IR photometry and
K. Belle and C. Mauche for comments on a 
draft of this manuscript. An anonymous referee provided a number of useful
comments leading to improvements in the presentation.
The research was supported by an AO-7 EUVE mini-grant and NASA grant NAG5-8644
to SBH and an AO-6 EUVE mini-grant to CRD. This work was supported in part by
the German DLR under grant 50 OR 9706 8.
 
\references

\reference{} Abbott, M., et al. 1996, ApJ Suppl., 107, 451

\reference{} Bowyer, S., \& Malina, R., 1991, in ``Extreme Ultraviolet 
Astronomy",
ed. R. Malina and S. Bowyer, (New York: Pergammon Press), 397

\reference{} Ciardi, D., Howell, S. B., Hauschildt, P., and Allard, F., 1998, 
ApJ, 504, 450.

\reference{} Cropper, M., 1990, Space Science Reviews, 54, 195. 

\reference{} Craig, N., et al. 1997, ApJ Suppl., 113, 131.

\reference{} Elias, J. H. et al. 1982, AJ, 87, 1824

\reference{} Fischer, A., \& Beuermann, K., 2001, A\&A, 373, 211.

\reference{} Frank, J., King, A., \& Raine, D., 1992, {\it Accretion Power in Astrophysics},
Cambridge University Press, New York.

\reference{} Glenn, J., Howell, S. S., Schmidt, G., Liebert, J., Grauer, A., \&
Wagner, R. M., 1994, ApJ., 424, 967.

\reference{} Howell, S. B., Gelino, D., \& Harrison, T., 2001, AJ, 121, 482.

\reference{} Howell, S. B., Ciardi, D., Dhillon, V., \& Skidmore, W., 
2000, ApJ, 530, 904.

\reference{} Howell, S. B., Sirk, M., Malina, R. F., Mittaz, J. P. D.,
\& Mason, K. O., 1995, ApJ, 439, 991

\reference{} Howell, S. B., Sirk, M., Ramsey, G., Cropper, M., Potter,
S., \& Rosen, S., 1997, ApJ, 485, 333

\reference{} Howell, S. B., Mitchell, K. J., \& Warnock, A. 1988, AJ, 95, 247

\reference{} Hurwitz, M., Sirk, M., Bowyer, S., \& Ko, Y., 1997, ApJ 477, 390

\reference{} King, A., 1995, in ``Cape Workshop on Magnetic Cataclysmic
Variables", eds. D. Buckley \& B. Warner, ASP Conf. Series Vol. 85, p. 21

\reference{} King, A., \& Watson, M., 1987, MNRAS, 227, 205

\reference{} Mauche, C., 1998, in ``Annapolis Workshop on Magnetic Cataclysmic 
Variables",
eds. C. Hellier \& K. Mukai, ASP Conf. Series Vol. 157, p. 157

\reference{} Paerels, F., Heise, J., \& van Teeseling, A., 1994, ApJ, 426, 313

\reference{} Paerels, F., Hur, M., Mauche, C., \& Heise, J., 1996, ApJ, 464, 884

\reference{} Ramsey, G., Mason, K., Cropper, M., Watson, M., \& Clayton, K., 
1994,
MNRAS, 270, 692

\reference{} Schwope, A., et al. 1998, in ``Wild Stars in the Old West", 
eds. S. B. Howell, E. Kuulkers, \& C. Woodward, ASP Conf. Series Vol. 137, p. 44

\reference{} Schwope, A., Schwarz, R., Sirk, M., \& Howell, S. B., 2001a, 
A\&A, 375, 419

\reference{} Schwope, A., Thomas, H.-C., Mantel, K.-H., \& Haefner, R., 2001b, 
A\&A, submitted.

\reference{} Sirk, M., et al. 1997, ApJ Suppl., 110, 347 

\reference{} Sirk, M., \& Howell, S. B., 1998, ApJ, 506, 824

\reference{} van Teeseling, A., Heise, J., \& Paerels, F., 1994, A\&A, 281, 119

\reference{} Warren, J., Sirk, M., \& Vallerga, J., 1995, ApJ, 445, 909.

\reference{} Warren, J., \& Mukai, K., 1996, in ``Astrophysics in the Extreme
Ultraviolet", eds. S. Bowyer \& R. Malina, Kluwer Publishers, p. 325

\reference{} Watson, M., King, A., Jones, D., \& Motch, C., 1989, MNRAS, 237,
299.

\reference{} Wickramasinghe D., \& Ferrario, L., 2000, PASP, 112, 873

\newpage

\begin{deluxetable}{lcc}
\tablewidth{4.5in}
\tablenum{1}
\tablecaption{Accretion Region Parameters for HU Aqr}
\tablehead{
\colhead{Parameter\tablenotemark{a}} &
\colhead{May 1996\tablenotemark{b}} &
\colhead{August 1998\tablenotemark{c}} \\
}
\startdata
Spot Radius (R$_{WD}$)       &   0.036    &    0.061(5) \\
Spot Radius (10$^7$ cm)      &   2.16     &    3.66 \\
Spot Area (10$^{15}$ cm$^2$) &   1.47     &    4.21 \\
Fractional Emitting Area     &   3.27$\times$10$^{-4}$  & 9.36$\times$10$^{-4}$ 
\\
Spot Height (R$_{WD}$)       &   0.021    &   0.023(3) \\
Spot Latitude (degrees)      &   34.5\tablenotemark{d}    &   36(7) \\
Spot Longitude (degrees)     &   49       &   49(1) \\
Accretion Region Temperature (K) & $\sim$124,000 & 240,000(40,000) 
\enddata
\tablenotetext{a}{We assume here a white dwarf radius of 6000 km 
(Ciardi et al. 1998).}
\tablenotetext{b}{Sirk \& Howell (1998)}
\tablenotetext{c}{Numbers in () are 1$\sigma$ errors.}
\tablenotetext{d}{This value was reported as 40 degrees in Sirk \& Howell 
(1988), 
but using the new binary inclination estimate of
85.6 degrees (Schwope et al. 2001a), compared with the older value of 
81, we subtract 4.6 degrees from 
the Sirk \& Howell value.}
\end{deluxetable}{}

\newpage

\begin{center}
Figure captions
\end{center}

Figure 1: August 1998 EUV light curve for HU Aqr. Various phases of interest
are marked on the plot.

Figure 2: Phase-resolved EUV spectra for HU Aqr. The four panels show (top to
bottom) the total summed dataset, the Bright phase, the Dip1 phase, and the Dip2 
phase. The 1$\sigma$ error is given for each panel 
and the final binned spectral resolution is 0.8\AA.
See Figure 1 for the phase intervals involved.

Figure 3: Orbitally phased J (top) and K (bottom) photometry of HU Aqr 
are plotted 
for the 1996 data (crosses) and the 1998 data (circles). The solid black
curve is the ellipsoidal model for the secondary star from Ciardi et al.
(1998) and has been overlayed upon both the 1996 and 1998 data. 
Estimated 1$\sigma$ uncertainties in both J and K are $\pm$0.2 mag.
Note the good agreement of the depth of the eclipse in the 1996 and 
1998 observations and how well
the ellipsoidal model fits the 1996 observations.

Figure 4: Historical retrospective of previous {\it EUVE} observations of HU
Aqr and the August 1998 high accretion state light curve (bottom panel).
Note the y-axis scale change for the August 1998 observation.
The dotted line is the global average {\it EUVE} light curve, scaled in each panel
to the maximum value. See text for details.

Figure 5: Eight consecutive binary orbits of HU Aqr from the 1998 {\it EUVE}
observation. 
Each light curve consists of $\sim$7400 seconds of data ($\sim$1 binary orbit) 
composed of four consecutive EUVE satellite orbits and phased on the binary
ephemeris. 
The data are summed in 20 sec bins. The top of each plot gives the start time for each
light curve. Note the rapid
changes that occur in each 2 hour time interval, especially near phases 0.7--0.9.

Figure 6: Summed spectrum for HU Aqr during low mass accretion states (1996, top) and 
a our high mass accretion spectrum (1998, bottom). 
The spectra are best fit with an absorbed
blackbody model yielding temperatures of $\sim$124,000 K and 240,000 K 
respectively. The dotted line indicates the 1$\sigma$ uncertainty in the flux,
the locations of expected and likely features are marked, 
and the final binned spectral resolution is 0.54\AA.

Figure 7: Plot of the phased 1998 J and K flux densities 
(F$_{1.2 \mu m}$ and F$_{2.2 \mu m}$) after the subtraction of
the secondary star mean level and ellipsoidal variations based on the model in
Ciardi et al. (1998). The vertical
lines and numbers indicate the divisions labeled and named in Figure 1.
The regions marked will be used again in Figures 8 \& 9.

Figure 8: Plot of the infrared and EUV flux densities. Each dataset 
has been normalized
such that the maximum point in each light curve is unity and the EUV data has been
re-sampled to match the J and K points respectively. The vertical lines
and numbers indicate the divisions labeled and named in Figure 1.

Figure 9: The J (F$_{1.2 \mu m}$) and K (F$_{2.2 \mu m}$) flux densities 
(as in Figure 7)
have been directly ratioed and plotted as a function of orbital phase.
The estimated uncertainty in the flux ratio is 1$\sigma$=$\pm$0.35.
The stellar eclipse points (region 5) have been omitted from this
figure and the J data have been re-sampled at the K data rate. The solid 
histogram is the EUV flux scaled such
that the maximum EUV point matches the maximum J/K flux ratio.
A F$_{1.2 \mu m}$/F$_{2.2 \mu m}$ 
ratio of 4.3 indicates that the IR emission is dominated by cyclotron radiation.
See text for details.


\newpage

\begin{figure}
\psfig{figure=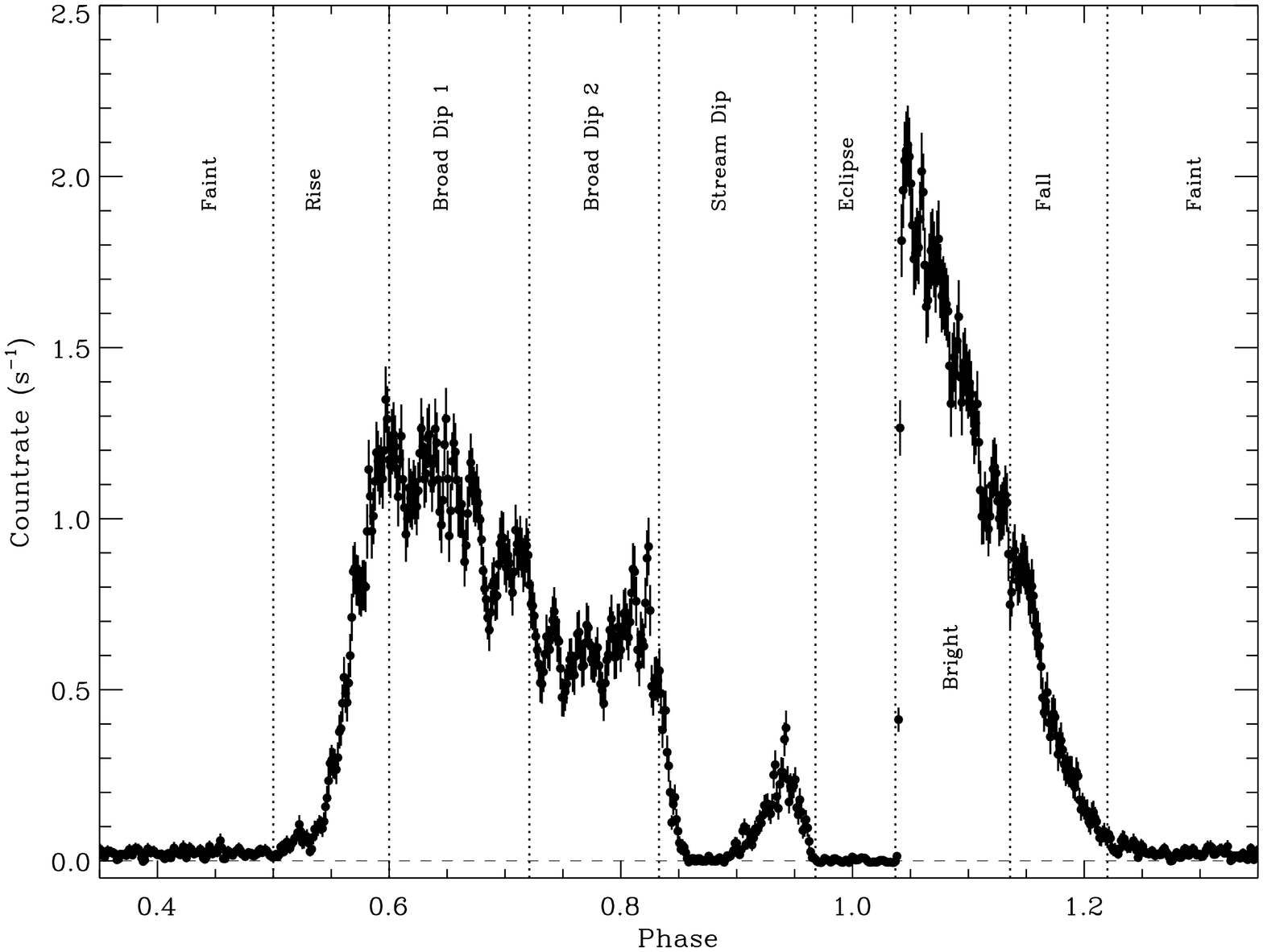,width=6in,angle=90}
\end{figure}

\begin{figure}
\psfig{figure=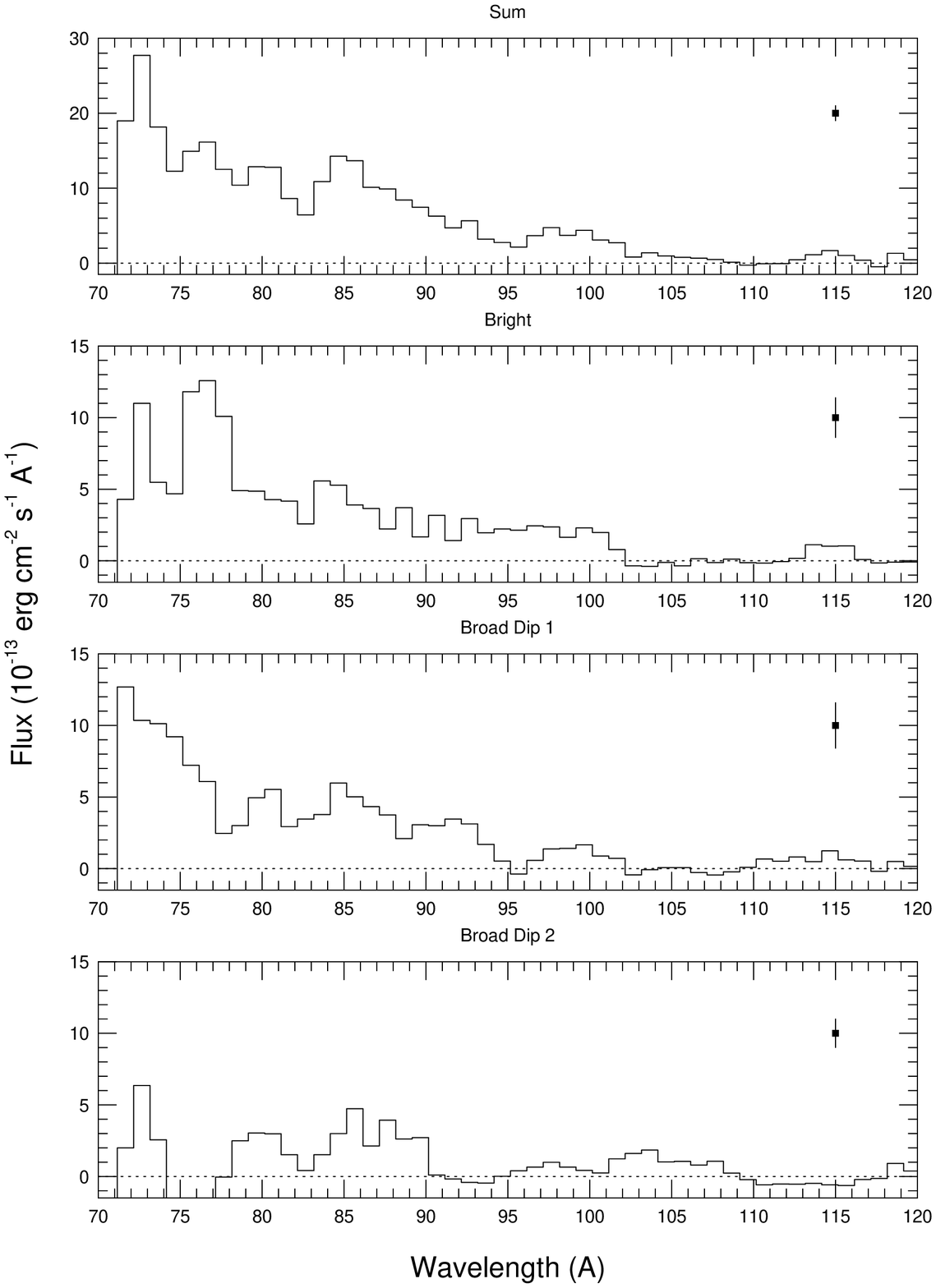,width=6in,angle=0}
\end{figure}

\begin{figure}
\psfig{figure=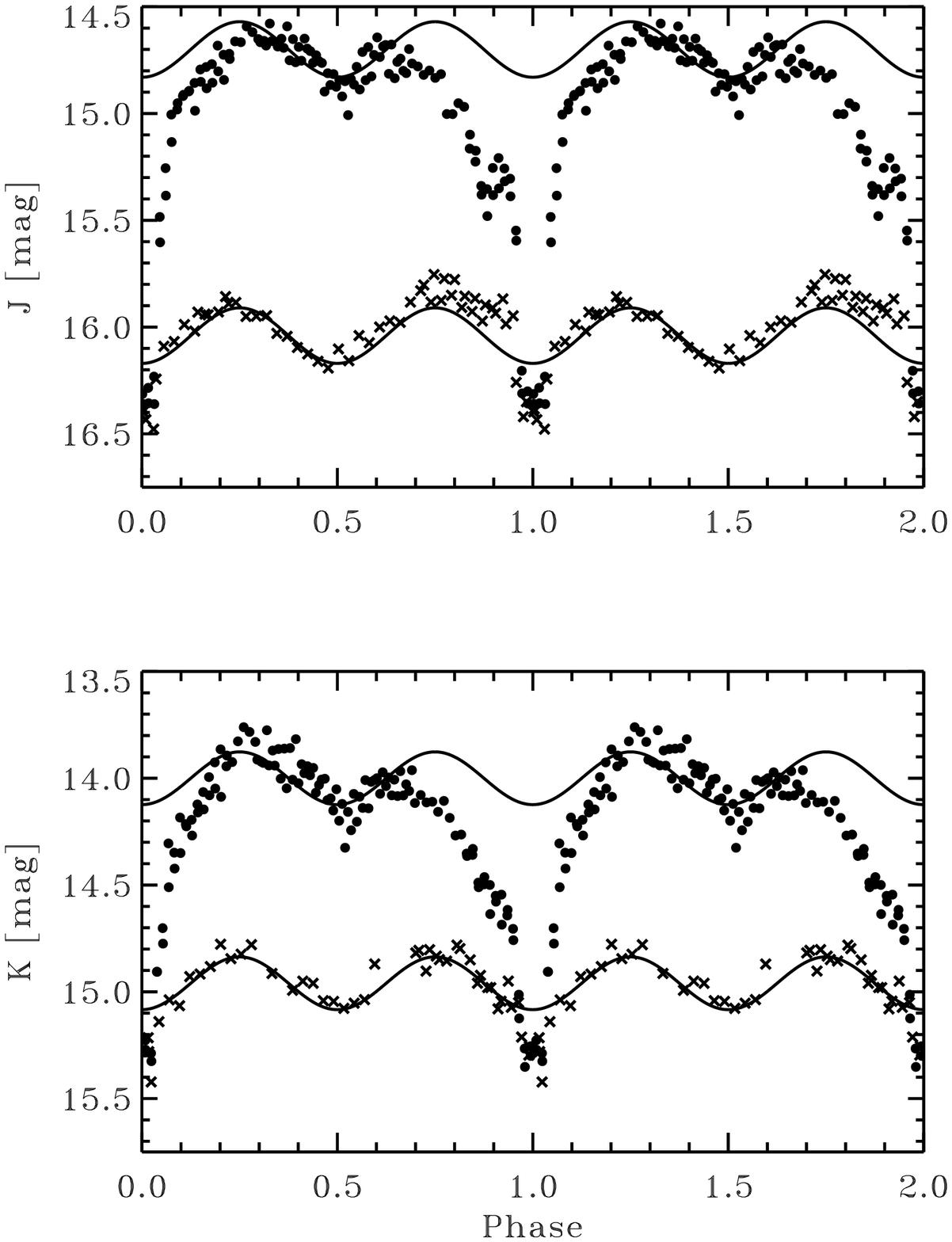,width=6in,angle=0}
\end{figure}

\begin{figure}
\psfig{figure=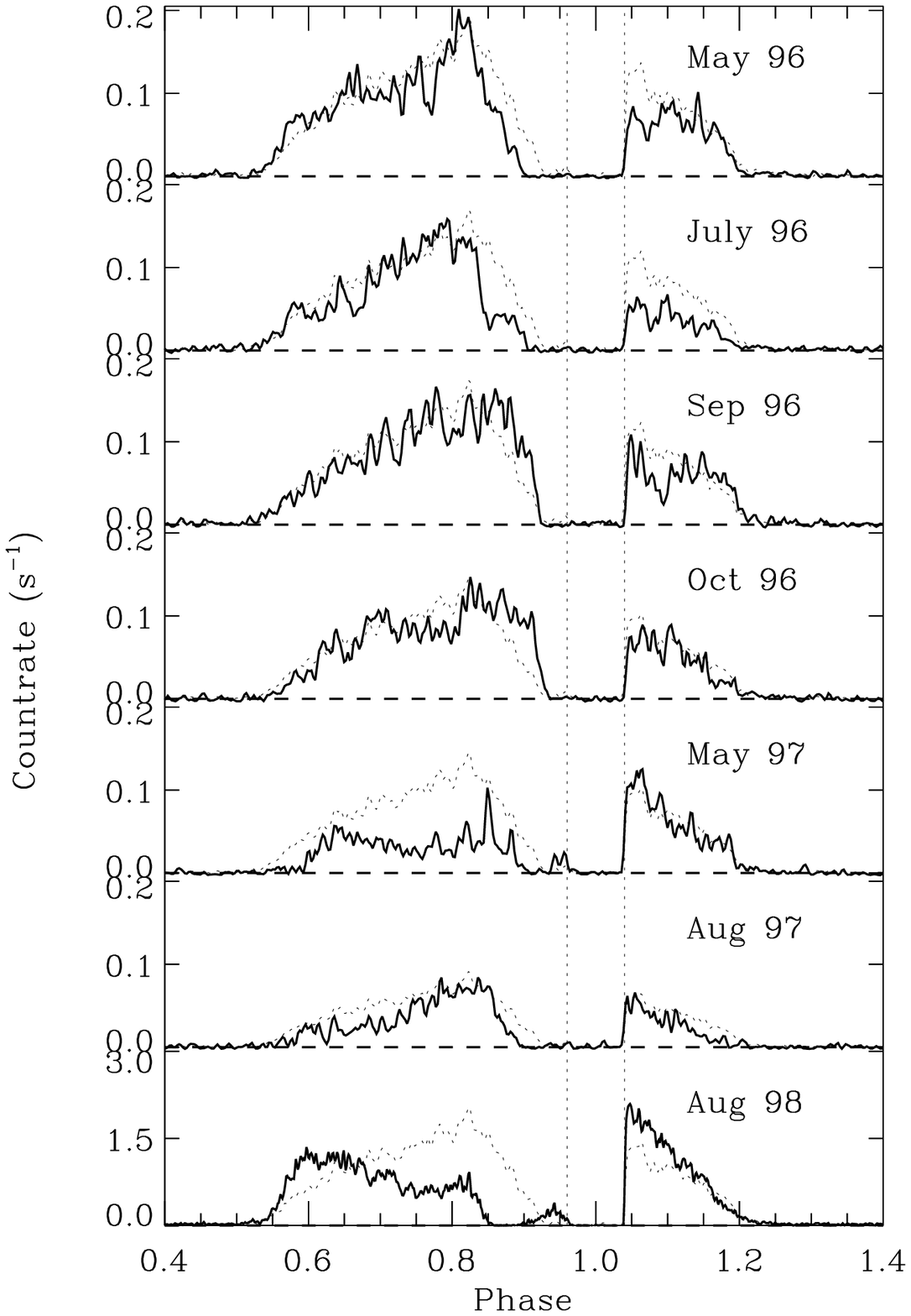,width=6in,angle=0}
\end{figure}

\begin{figure}
\psfig{figure=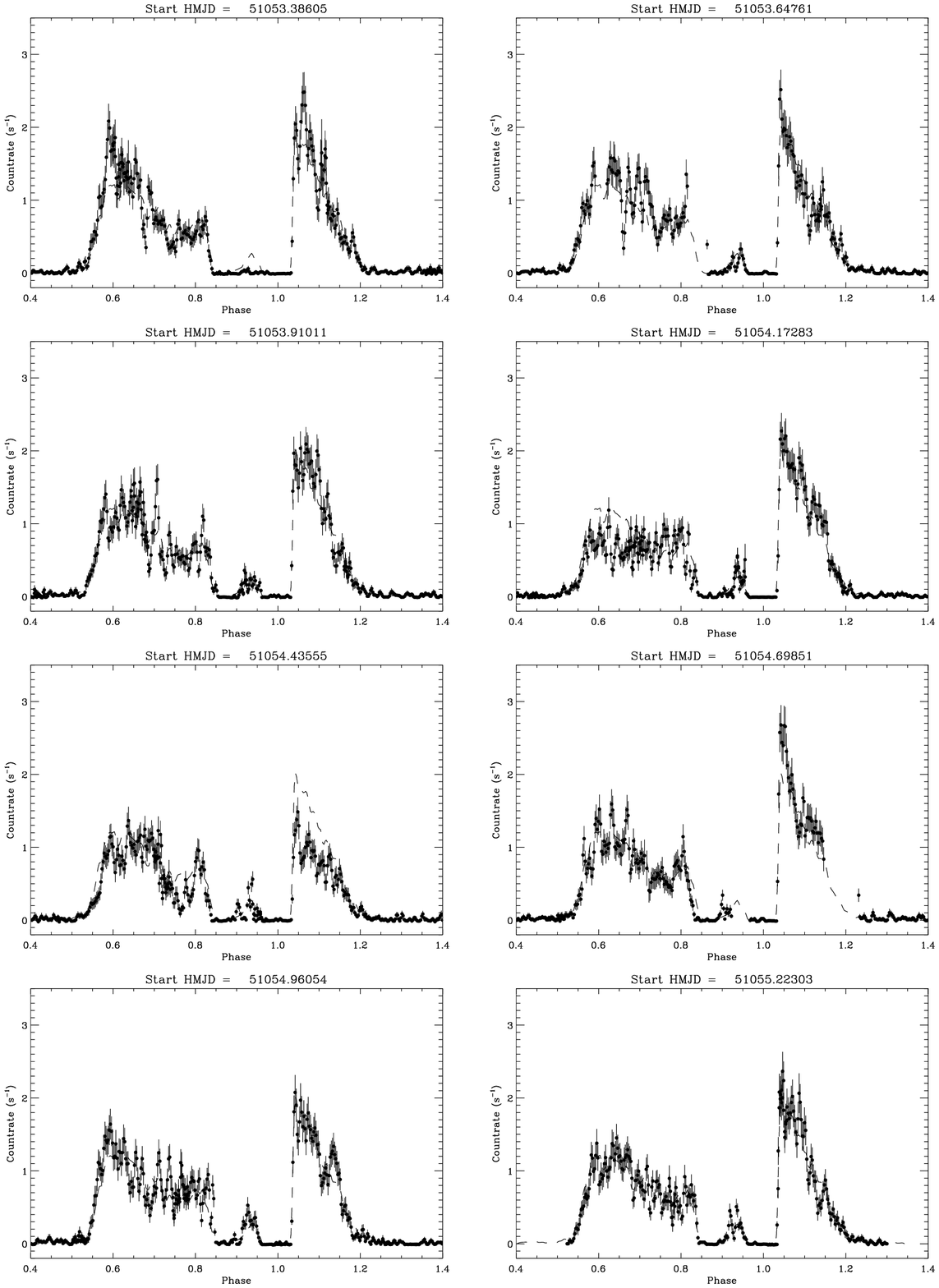,width=6in,angle=0}
\end{figure}

\begin{figure}
\psfig{figure=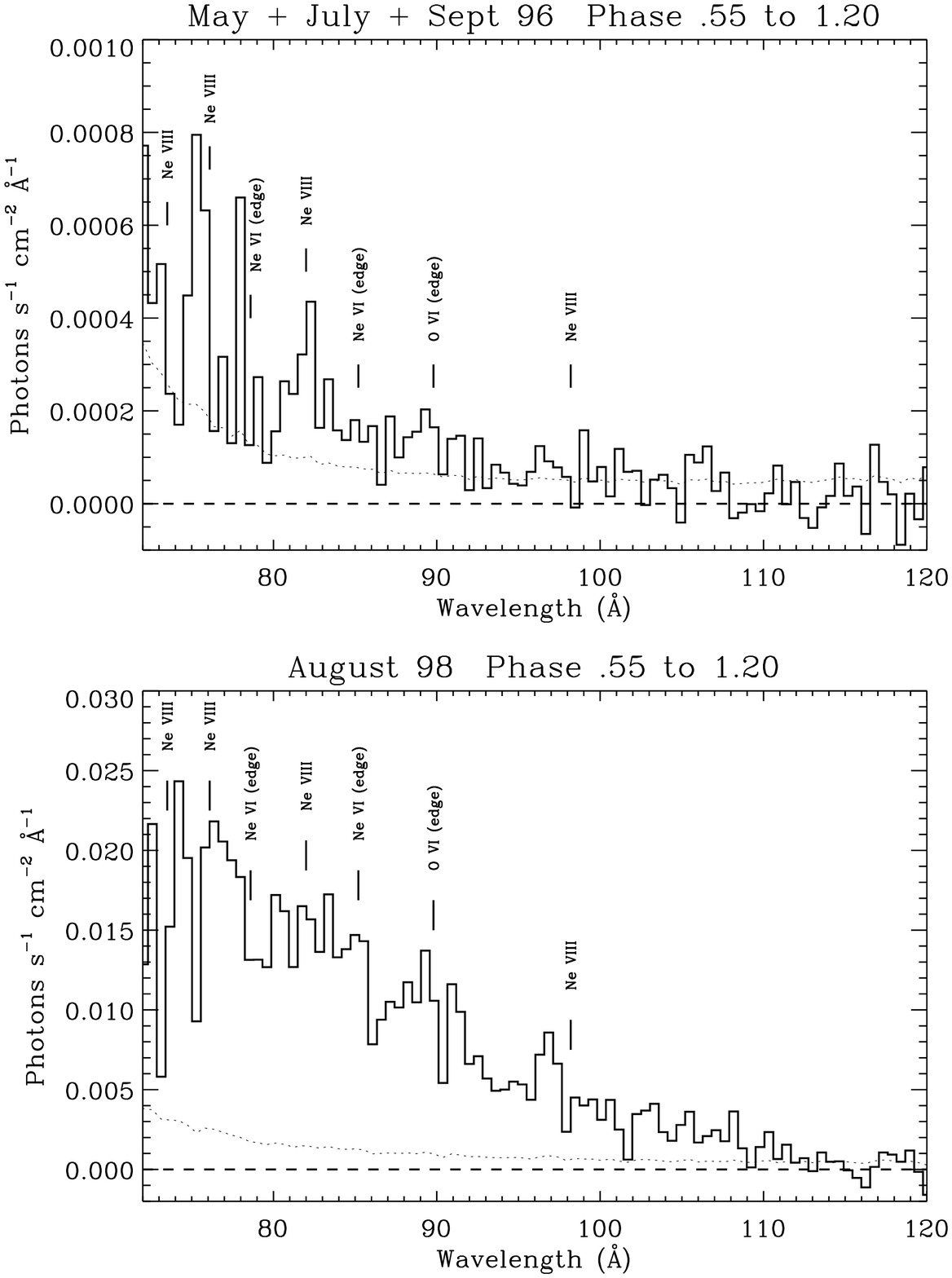,width=6in,angle=0}
\end{figure}

\begin{figure}
\psfig{figure=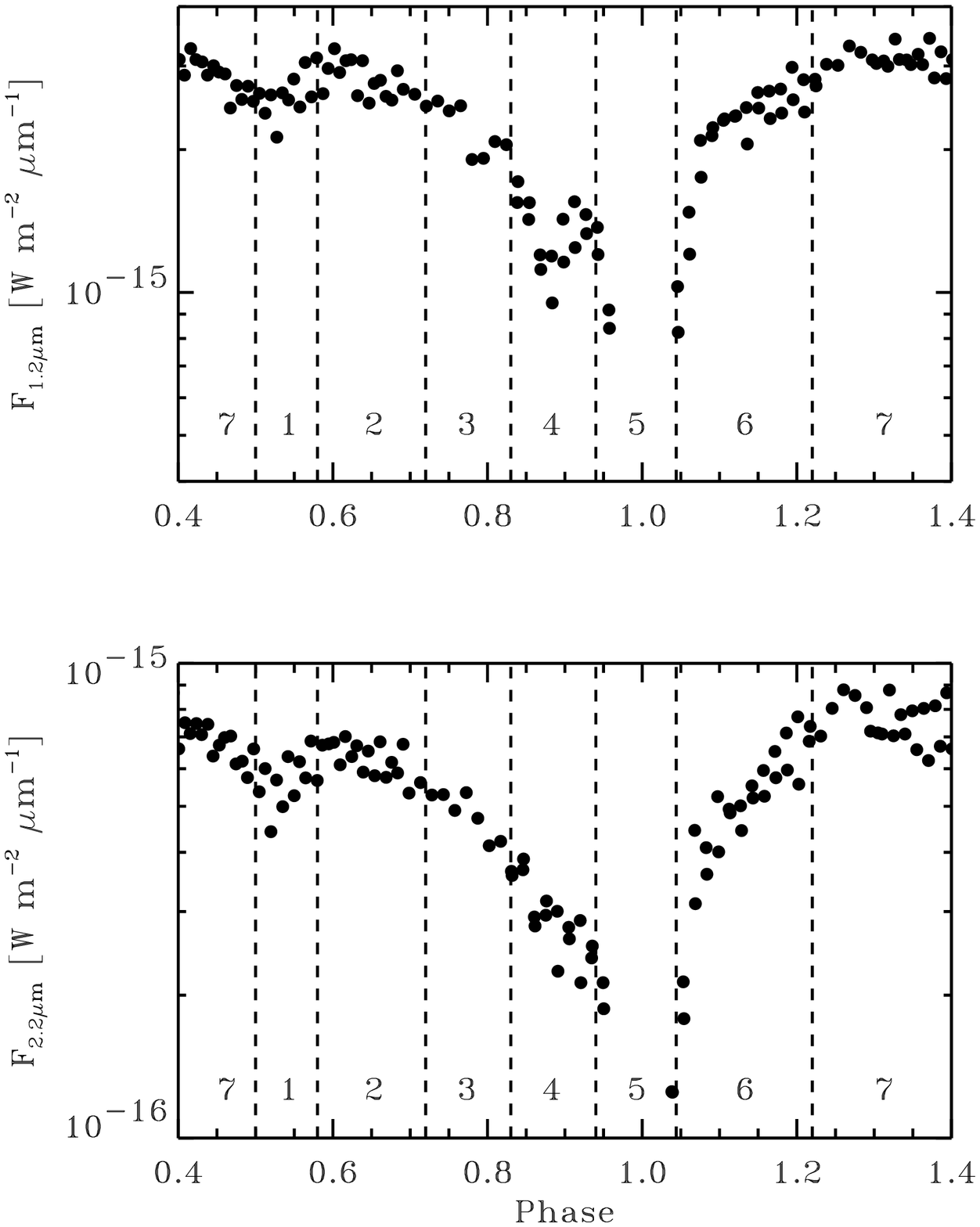,width=6in,angle=0}
\end{figure}

\begin{figure}
\psfig{figure=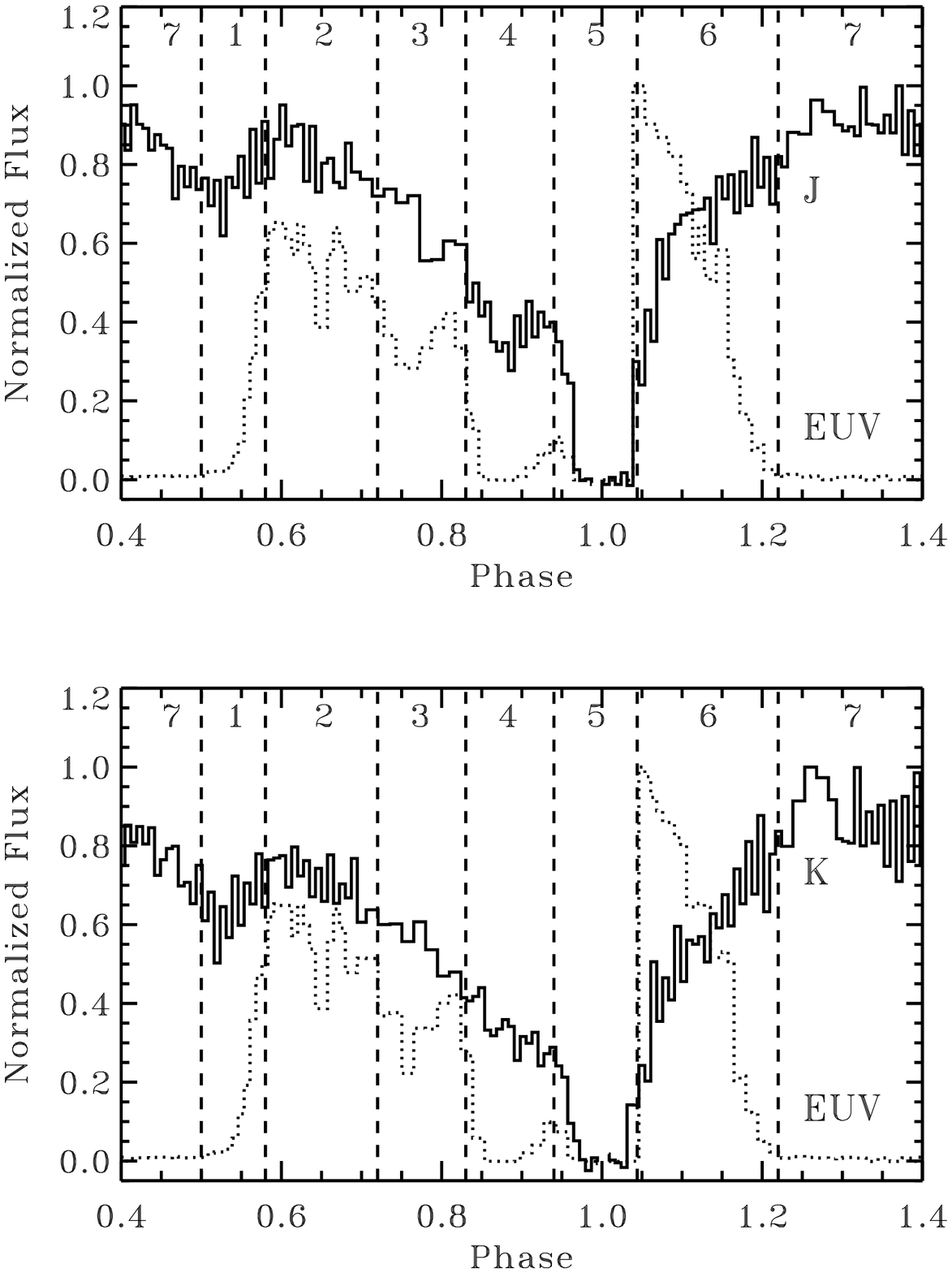,width=6in,angle=0}
\end{figure}

\begin{figure}
\psfig{figure=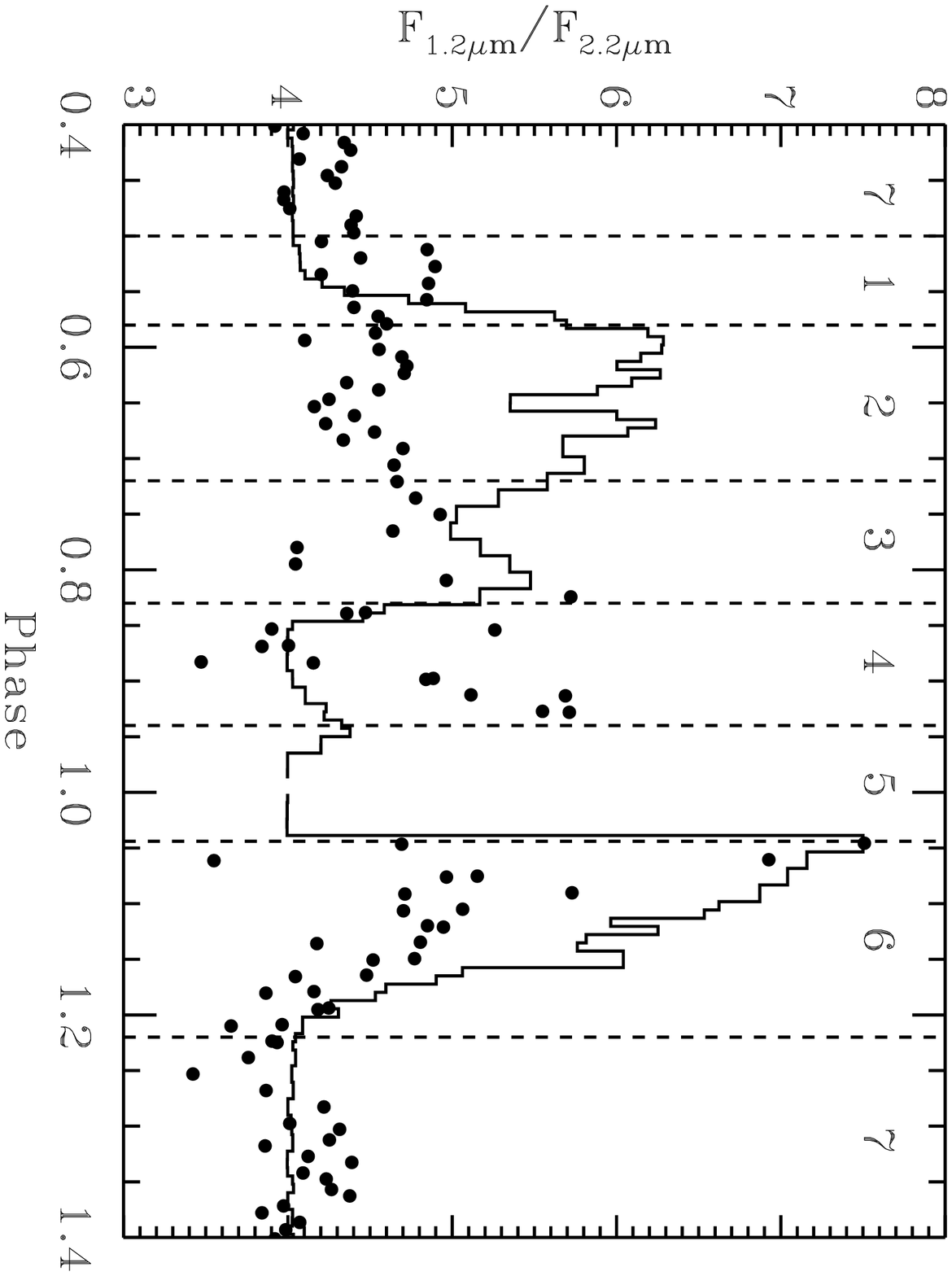,width=6in,angle=0}
\end{figure}

\end{document}